\documentclass[11pt]{article}

\usepackage{graphicx}

\newcommand{\be}{\begin{equation}}
\newcommand{\ee}{\end{equation}}

\newcommand{\p}{\partial}

\newcommand{\al}{\alpha}

\begin{document}

\title{\vskip-2.5cm The gravitational two-body potential \\ generated by
                                binary  % stars
                                        systems}   
\author{{\bf Renato Spigler$^{1,2,*}$}	  \\
   $^1$Department of Mathematics and Physics, Roma Tre University   \\
             1, Largo S. Leonardo Murialdo, 00146 Roma, Italy    \\
     $^2$Institute for Complex Systems ISC-CNR \\
       $^*${\em Project IGNITOR Group}   \\
                    {\tt spigler@mat.uniroma3.it} }

%\date{}

\maketitle    

\begin{abstract}
\noindent We evaluate the {\em three-dimensional}, {\em non-axis-symmetric},
%%{\em non-axisymmetric}
{\em time-dependent} Newton potential generated by a pair of mutually orbiting
objects such as pairs of ordinary or neutron stars and, in some approximations,
black holes, spinning around each other. The `vertical component' of the
gravitational force (that is, that orthogonal to the plane of their orbit)
%(the disc, in case of black holes)
is also evaluated, along with the other components of the field. The
pseudo-Newtonian Paczy\'{n}ski-Wiita form of the potential is also computed.
The effect of the asymmetry due to the more common case of different masses is
stressed.
\vskip0.4cm
%%%
%%% PACS:
%%%   02.30.Em Potential theory
%%%   95.10.Ce Celestial mechanics
%%%   97.60.Jd Neutron stars
%%%   97.60.Lf Black holes
%%%   97.80.−d Binary and multiple stars
%%%
%%% 2020 Mathematics Subject Classification:
%%%    	31-08 Computational methods for problems pertaining to potential theory 
%%%  	31Cxx Generalizations of potential theory
%%% 	65Exx Numerical methods in complex analysis (potential theory, etc.)
%%% 	70F15 Celestial mechanics
%%% --- this drafi is now  arXiv:submit/4883240 [math-ph] 8 May 2023
%%%
\end{abstract}

\section{Introduction}
 
  Consider a system of two bodies, such as two ordinary stars or neutron stars,
or even black holes (with some limitations as seen below), orbiting around each
other. These bodies may have have different masses, say $m_1$ and $m_2$ ($m_1
\geq m_2$. e.g.), while their sizes are much smaller than their distance, so
that they can be considered as concentrated at two points. For simplicity, we
assume first that they perform a {\em circular} orbit, each spinning around
their common barycenter, with radii $d_1 > 0$, $d_2 > 0$, being $d = d_1 + d_2$
their distance. Under such conditions,
the gravitational (Newton) potential they generate is three-dimensional,
{\em not} axis-symmetric, and time-dependent.

\vskip0.1cm

  It should recalled that nowadays binary systems are found very frequently,
and the remarkable fact that the gravitational potential generated by them is
{\em time dependent} (in contrast to that due to a single body) can be observed
even when a small body orbits around a much more massive one, e.g., a
``shepherd planet'', spinning around a black hole or a neutron star, which would
generate
a modulated gravitational field. This makes it even more frequent the
occurrence of the effects described in this paper.

  We will also consider the more general case when the two bodies orbit around
each other following {\em elliptic} paths, with their barycenter located at one
of the focii.

  We consider an equilibrium state, when gravitational attraction is balanced
by centrifugal force.
  We adopt a reference system of cylindrical coordinates, $(R,\varphi,z)$, and
locate initially the two bodies of masses $m_1$, $m_2$, at the points $P_1 =
(-d_1,0,0)$, $P_2 = (d_2,0,0)$, in rectangular coordinates, while the center of
gravity is located at the origin. Therefore, the motion of $P_1$ and $P_2$ is
described by
$$
   x_1 = d_1 \cos \Omega t, \ \  y_1 = d_1 \sin \Omega t, \ \  z_1 = 0,
$$
$$
   x_2 = d_2 \cos(\Omega t + \pi), \ \  y_2 = d_2 \sin(\Omega t + \pi),
      \ \  z_2 = 0,
$$
where $t$ denotes time, $\Omega$ is the orbiting frequency, and we used the
fact that the two bodies remain diametrally opposite during their motion. Hence
the distances of the generic point $P = (x,y,z) \equiv (R,\varphi,z)$ from
$P_1$ and $P_2$ are given by
$$
     \overline{P_1P}^2 = (R \cos \varphi - d_1 \cos \Omega t)^2
        + (R \sin \varphi - d_1 \sin \Omega t)^2 + z^2,
$$    
$$
     \overline{P_2P}^2 = (R \cos \varphi - d_2 \cos(\Omega t + \pi))^2
        + (R \sin \varphi - d_2 \sin(\Omega t + \pi))^2 + z^2,
$$    
  Note that $r \equiv \overline{OP} = \sqrt{R^2 + z^2}$, and there is no much
difference between $r$ and $R$ as long as $z \ll R$ (or $z \ll r =
\overline{OP}$). An advantage of using the cylinder coordinate $R$ instead of
the polar coordinate $r$ (as in \cite{Coppi-PhysLett2018,Coppi-Rep }) is that
$R$ is independent of $z$, and hence differentiating the potential with respect
to $z$ in \S~3 will be simpler.

\section{The gravitational potential}

  The gravitational (Newton) potential will be
$$
    \Phi \equiv \Phi(R,\phi,z;t) = - \frac{G m_1}{\overline{P_1P}}
              - \frac{G m_2}{\overline{P_2P}}
$$
$$
  = \frac{-G m_1}{\sqrt{R^2 + d_1^2 - 2 R d_1 \cos(\varphi - \Omega t) + z^2}}
 + \frac{-G m_2}{\sqrt{{R^2 + d_2^2 + 2 R d_2 \cos(\varphi - \Omega t) + z^2}}}
$$
$$
   = \frac{-G m_1}{\sqrt{R^2 + d_1^2 + z^2}} \frac{1}{\sqrt{1 - \frac{2 R d_1}
      {R^2 + d_1^2 + z^2} \, \cos(\varphi - \Omega t)}}
$$
\be
    + \frac{-G m_2}{\sqrt{R^2 + d_2^2 + z^2}} \frac{1}{\sqrt{1 + \frac{2 R d_2}
        {R^2 + d_2^2 + z^2} \, \cos(\varphi - \Omega t)}}.
                                                         \label{gravit-pot}
\ee 
Note that $\varphi - \Omega t$ denotes a propagation along $\varphi$, i.e.,
azimuthal.
 Here $G$ denotes the gravitational constant, and we will assume that $d \ll R$,
hence $d_{\ell} \ll R$, $\ell=1,2$, and $z \ll R$. Recall that, instead, $r =
\overline{OP} = \sqrt{R^2 + z^2}$.

  We set, for short,
\be
   a_{\ell} := \frac{G m_{\ell}}{\sqrt{R^2 + d_{\ell}^2 + z^2}}
            = \frac{G m_{\ell}}{D_{\ell}^{1/2}},
        \quad
  \alpha_{\ell} := \frac{2 R d_{\ell}}{R^2 + d_{\ell}^2 + z^2}
                = 2 \, \frac{d_{\ell} R}{D_{\ell}},    
                                                            \label{a-al-ell}
\ee
for $\ell=1,2$, where
\be
       D_{\ell} := R^2 + d_{\ell}^2 + z^2,  \ \  \ell =1,2,          \label{Di}
\ee
and $\theta := \varphi - \Omega t$. Note that $0 < \alpha_{\ell} \leq 1$ for
every $z \neq 0$ and $d_{\ell}/R \leq 1$, but $0 < \alpha_{\ell} < 1$ for every
$z \neq 0$ and $d_{\ell}/R < 1$.
Note also that $a_{\ell} = {\cal O}\left(\frac{1}{R}\right)$, and $\alpha_i
\approx 2 \frac{d_{\ell}}{R}$. Thus we can write concisely
\be
   -\Phi = \frac{a_1}{\sqrt{1 - \alpha_1 \, \cos \theta}}
          + \frac{a_2}{\sqrt{1 + \alpha_2 \, \cos \theta}},
                                                           \label{potential}
\ee
and also
\be
   -\Phi = a_1 \, \left[\frac{c_0^{(1)}}{2}
               + \sum_{n=1}^{\infty} c_n^{(1)} \cos n \theta\right]
         + a_2 \, \left[\frac{c_0^{(2)}}{2}
               + \sum_{n=1}^{\infty} c_n^{(2)} \cos n \theta\right].
                                                             \label{pot-1-2}
\ee
  This function can be evaluated in several ways, resorting for instance to
elliptic integrals or to binomial series.

\vskip0.1cm

\subsection{Using elliptic integrals}

  There is an elegant way to express the potentials we are interested in, by
means of {\em elliptic integrals}. The potential in (\ref{potential}) is an
even and $2 \pi/\Omega$-{\em periodic} function, and hence it seems natural to
expand it in Fourier series (of cosines only). However, evaluating its Fourier
coefficients leads to handling elliptic integrals \cite[Ch.~17]{Abramowitz},
\cite[Ch.~19]{DLMF}, \cite{elliptic}. In fact, we have (for either $\ell=1$ or
$\ell=2$)
$$
   \frac{1}{\sqrt{1 - \alpha_{\ell} \, \cos \theta}} = \frac{c_0^{(\ell)}}{2}
         + \sum_{n=1}^{\infty} c_n^{(\ell)} \, \cos n \theta,    \quad \ell=1,2,
$$
where
\be
     c_n^{(\ell)} = \frac{2}{\pi} \int_0^{\pi} \frac{\cos n \theta}
        {\sqrt{1 - \alpha_{\ell} \, \cos \theta}} \, d \theta,  \ \
                                      n = 0,1,2,\ldots,    \label{Fourier-n}
\ee
and in particular the mean value $\frac{c_0^{(\ell)}}{2}$ is given by
\be
   c_0^{(\ell)} = \frac{1}{\pi} \int_0^{\pi} \frac{1}
    {\sqrt{1 - \alpha_{\ell} \, \cos \theta}} \, d \theta.   \label{Fourier-0}
\ee
The latter leads to the time-independent term in the representation of the
potential $\Phi$, and can be obtained in terms of {\em complete elliptic
integral of the first kind} \cite{elliptic}. Indeed,
$$
     -\langle \Phi \rangle = \frac{a_1}{2 \pi}
   \int_{-\pi}^{\pi} \frac{d \theta}{\sqrt{1 - \alpha_1 \cos \theta}}
         + \frac{a_2}{2 \pi}
     \int_{-\pi}^{\pi} \frac{d \theta}{\sqrt{1 + \alpha_2 \cos \theta}}
$$
\be
    = \frac{a_1}{\pi} I_0(\alpha_1) + \frac{a_2}{\pi} I_0(\alpha_2), 
                                                         \label{ave-phi}
\ee
since
$$
    \int_{-\pi}^{\pi} \frac{d \theta}{\sqrt{1 + \alpha_2 \cos \theta}}
       = 2 \int_0^{\pi} \frac{d \theta}{\sqrt{1 + \alpha_2 \cos \theta}}
       = 2 \int_0^{\pi} \frac{d \delta}{\sqrt{1 - \alpha_2 \cos \delta}},
$$
having set $\theta = \pi - \delta$, and defined
$$
   I_0(\alpha_1) :=
       \int_0^{\pi} \frac{d \theta}{\sqrt{1 - \alpha_1 \cos \theta}} =
 \frac{2}{\sqrt{1+\alpha_1}} \int_0^{\pi/2} \frac{dx}{\sqrt{1 - k_1^2 \sin^2 x}},
$$
where we set $\theta = \pi - 2x$ and
\be
           k_1^2 := \frac{2 \alpha_1}{1 + \alpha_1}.              \label{k1-2}
\ee
Note that $k_1^2 < 1$, being $0 < \alpha_1 < 1$ (assuming $d_1/R < 1$).
Therefore, we can write
$$
     I_0(\alpha_1) = \frac{2}{\sqrt{1 + \alpha_1}}
       \int_0^{\pi/2} \frac{dx}{\sqrt{1 - k_1^2\sin^2 x}}
          = \frac{2}{\sqrt{1+\alpha_1}} \, K(k_1), 
$$
introducing the {\em complete elliptic integral of the first kind}
\cite[Ch.~17]{Abramowitz}, \cite[Ch.~19]{DLMF}, \cite{elliptic},
\be
    K(k_1) := \int_0^{\pi/2} \frac{d x}{\sqrt{1 - k_1^2 \sin^2 x}}
      = \int_0^1 \frac{d t}{\sqrt{1 - t^2} \sqrt{1 - k_1^2 t^2}}
                                                             \label{first-k}
\ee
Note that, changing $\theta$ into $\pi - \theta$ in the second integral in
(\ref{ave-phi}), we have seen that $I_0(-\alpha_2) = I_0(\alpha_2)$. Therefore,
defining $k_2^2$ as in (\ref{k1-2}), just replacing the index 1 with 2, we have
$$
  -\langle \Phi \rangle = \frac{2 a_1}{\pi \sqrt{1 + \alpha_1}}
         \int_0^{\pi/2} \frac{dx}{\sqrt{1 - k_1^2 \sin^2 x}}
      + \frac{2 a_2}{\pi \sqrt{1 + \alpha_2}}
          \int_0^{\pi/2} \frac{dx}{\sqrt{1 - k_2^2 \sin^2 x}}
$$
$$
  = \frac{2 a_1}{\pi \sqrt{1 + \alpha_1}}
         \int_0^1 \frac{dt}{\sqrt{1 - k_1^2 t^2} \, \sqrt{1 - t^2}}
      + \frac{2 a_2}{\pi \sqrt{1 + \alpha_2}}
          \int_0^1 \frac{dt}{\sqrt{1 - k_2^2 t^2} \, \sqrt{1 - t^2}}
$$
\be
     = \frac{2 a_1}{\pi \sqrt{1 + \alpha_1}} \, K(k_1)
           + \frac{2 a_2}{\pi \sqrt{1 + \alpha_2}} \, K(k_2), \ \
                 \alpha_{\ell} = k_{\ell}^2, \ \ell=1,2.
                                                          \label{ave-Phi}   
\ee

  It is noteworthy that the function $K(k)$ can be computed as a power series
involving Legendre polynomials $P_n(x)$ \cite{Legendre},
\be
   K(k) =
 \frac{\pi}{2} \sum_{n=0}^{\infty} \left(\frac{(2n)!}{2^{2n} (n!)^2}\right)^2 k^{2n}
           = \frac{\pi}{2} \sum_{n=0}^{\infty} (P_{2n}(0))^2 k^{2n}.
                                                        \label{K-with-Leg}
\ee
Thus, being
$$
     \frac{(2n)!}{2^{2n} (n!)^2} = \frac{(2n-1)!!}{(2n)!!},
$$
we have more explicitly
\be
   K(k) = \frac{\pi}{2} \left[1 + \left(\frac{1}{2}\right)^2 k^2  
         + \left(\frac{1 \cdot 3}{2 \cdot 4}\right)^2 k^4 + \ldots
       + \left(\frac{(2n-1)!!}{(2n)!!}\right)^2 k^{2n} + \ldots \right].
                                                       \label{K-Leg-simple}
\ee
  Recall that the {\em generating function} of the Legendre polynomials is
$$
     \frac{1}{\sqrt{1 - 2 x t + t^2}} = \sum_{n=0}^{\infty} P_n(x) \, t^n,
$$
so that, e.g.,
$$
      \frac{1}{\sqrt{1 - x}} = \sqrt{2} \sum_{n=0}^{\infty} P_n(x).
$$

 It is also remarkable that the function $K(k)$ might be approximated fast and
very efficiently by the so-called {\em arithmetic-geometric mean}, agm$(x,y)$,
\cite{arithmetic} as
\be
   K(k) = \frac{\frac{\pi}{2}}{\mbox{agm}\!\left(1,\sqrt{1-k^2} \right)}
                                                             \label{K-agm}
\ee
which allows a very simple and efficient computation of it
\cite[\S~19.8]{Carlson}, besides being representable by means of the Gauss'
hypergeometric function as
\be
      K(k) = \frac{\pi}{2} \, _2F_1(1/2,1/2,1;k^2)        \label{K-hypergeo}
\ee
\vskip0.2cm

REcall that the arithmetic-geometric mean of two real numebrs, $x$ and $y$,
is defined as follows. Set $a_0 := x$ and $g_0 := y$, and define recursively
the two independent sequences $\{a_n\}$ and $\{g_n\}$ by 
\be
    a_{n+1} := \frac{a_n + g_n}{2},  \quad  g_{n+1} := \sqrt{a_n g_n}.
                                                         \label{seq-for-def}
\ee
It can be shown that both sequences $\{a_n\}$ and $\{g_n\}$ converge to the
same real number, denoted by $M(x,y)$, or AGM$(x,y)$, or agm$(x,y)$
\be
     \mbox{agm}(x,y) := \lim_{n \to \infty} a_n = \lim_{n \to \infty} g_n,
                                                             \label{def-agm}
\ee
and it is known that the convergence is very fast

\vskip0.2cm

  Computing the other Fourier coefficients in terms of elliptic integrals is
less simple. We have 
\be
    c_n^{(\ell)} = \frac{2}{\pi} \int_0^{\pi} \frac{\cos n \theta}
                   {\sqrt{1 - \alpha_{\ell} \cos \theta}} \, d \theta,
\ee
and thus, for $n=1$,
\be
   c_1^{(\ell)}
         = \frac{2}{\pi} \int_0^{\pi} \frac{\cos \theta}
               {\sqrt{1 - \alpha_{\ell} \cos \theta}} \, d \theta,
\ee
\be
   c_2^{(\ell)}
         = \frac{2}{\pi} \int_0^{\pi} \frac{\cos 2 \theta}
               {\sqrt{1 - \alpha_{\ell} \cos \theta}} \, d \theta.
\ee

\vskip0.1cm

\subsection{Using binomial series}   

Being $|\alpha_{\ell} \cos \theta| \leq \alpha_{\ell} < 1$ (indeed $\ll 1$), we
can instead expand both terms in (\ref{potential}) in convergent binomial
series, obtaining
$$
   (1 + \alpha_2 \cos \theta)^{-1/2} 
      = \sum_{k=0}^{\infty} {-1/2 \choose k} \alpha_2^k \cos^k \theta,
$$
and similarly for the other term, hence
$$
  -\Phi =
  a_1 \, \sum_{k=0}^{\infty} {-1/2 \choose k} (-1)^k \alpha_1^k \cos^k \theta
    + a_2 \, \sum_{k=0}^{\infty} {-1/2 \choose k} \alpha_2^k \cos^k \theta
$$
\be
   = \sum_{k=0}^{\infty} {-1/2 \choose k} \left[a_1 (-1)^k \alpha_1^k 
  + a_2 \alpha_2^k\right] \, \cos^k (\varphi - \Omega t).    \label{asymm-Phi}
\ee
  Note that $\cos^k \theta$ can be expressed as a linear combination of
$\cos(j \theta)$, for $j=0,1,\ldots,k$ (this can be done using Chebyshev
polynomials of the first kind \cite{Chebyshev}), and this is important since we
are interested, rather, in evaluating the coefficient of $\cos(k \theta)$, for
each $k$, in (\ref{asymm-Phi}).
  Note that, in general, {\em all harmonics} appear in (\ref{asymm-Phi}). This
is due to the lack of symmetry coming from having different masses. In fact, if
$m_1 = m_2$, we have $d_1 = d_2 = d/2$,
$$
    a_1 = a_2 = \frac{G m_1}{\sqrt{R^2 + \frac{d_1^2}{4} + z^2}}
                     \approx \frac{G m_1}{R},
$$
$$
   \alpha_1 = \alpha_2 = \frac{R d_1}{R^2 + \frac{d_1^2/4 + z^2}{R^2}}
                         \approx \frac{d_1}{R},
$$
and thus
\be
    -\Phi 
  = 2 a_1 \sum_{j=0}^{\infty} {-1/2 \choose 2j} \alpha_1^{2j} \, 
                 \cos^{2j} (\varphi - \Omega t),             \label{symm-Phi}
\ee
and only {\em even harmonics} appear. Hereafter we will write $\theta :=
\varphi - \Omega t$, for short.

\vskip0.1cm

 Incidentally, if we would consider two bodies rotating on {\em elliptic orbits}
around their common barycenter, located at one of the focii, the additional
lack of symmetry would let appear also sine functions.
  
  We first separate in (\ref{asymm-Phi}) the even from the odd harmonics:
$$
  -\Phi = \sum_{j=0}^{\infty} {-1/2 \choose 2j} \left[a_2 \alpha_2^{2j} 
         + a_1 \alpha_1^{2j} \right] \, \cos^{2j} \theta 
$$
$$
   + \sum_{j=0}^{\infty} {-1/2 \choose 2j+1} \left[a_2 \alpha_2^{2j+1} 
         - a_1 \alpha_1^{2j+1} \right] \, \cos^{2j+1} \theta 
$$
\be
    = \sum_{j=0}^{\infty} A_j
     + \sum_{j=1}^{\infty} \sum_{i=0}^{j-1} B_{ji} \, \cos((2j-2i)\theta)
     + \sum_{j=0}^{\infty} \sum_{i=0}^j C_{ji} \, \cos((2j+1-2i)\theta),
                                                             \label{Phi-ABC}
\ee
where
\be
    A_j := {-1/2 \choose 2j} \frac{1}{2^{2j}} {2j \choose j}
      \left[a_2 \alpha_2^{2j} + a_1 \alpha_1^{2j} \right], \ \  j=0,1,2,\ldots,
                                                                 \label{Aj}
\ee
\be
    B_{ji} := {-1/2 \choose 2j} \frac{2}{2^{2j}} {2j \choose i}
      \left[a_2 \alpha_2^{2j} + a_1 \alpha_1^{2j} \right],  \ \  j=1,2,\ldots,
                                            \, i=0,1,\ldots,   \label{Bji}
\ee
\be
    C_{ji} := {-1/2 \choose 2j+1} \frac{1}{2^{2j}}
       {2j+1 \choose i} \left[a_2 \alpha_2^{2j+1} - a_1 \alpha_1^{2j+1} \right],
                      \ \  j,i=0,1,2,\ldots,                    \label{Cji}
\ee
and we used the trigonometric formulae relating even and odd powers of cosines
to cosines of multiples of their arguments \cite{cos-powers},
$$
    \cos^k \theta = \frac{2}{2^k} \sum_{i=0}^{\frac{k-1}{2}} {k \choose i}
        \cos((k-2i)\theta), \ \ \mbox{if} \ k \ \mbox{is odd},
$$
$$
    \cos^k \theta = \frac{1}{2^k} {k \choose \frac{k}{2}}
        + \frac{2}{2^k} \sum_{i=0}^{\frac{k}{2}-1} {k \choose i}
          \cos((k-2i)\theta), \ \ \mbox{if} \ k \ \mbox{is even}, 
$$
but in the latter we intend that for $k=0$ we have $\cos^0 \theta = 1$. In
these formulae we could recognize the Chebyshev polynomials of the first kind,
$T_n(x)$, defined by
\be
        T_n(\cos \theta) := \cos(n \theta), \ \ n=0,1,2,\ldots
                                                          \label{Cebyshev}
\ee

In order to single out the {\em full contribution} (the coefficient) {\em of
each harmonic}, we should interchange the order of summation in
(\ref{Phi-ABC}), obtaining
\be
 -\Phi = \sum_{j=0}^{\infty} A_j + \sum_{s=1}^{\infty} \tilde{B}_s \, \cos(2s \theta)
            + \sum_{s=0}^{\infty} \tilde{C}_s \, \cos((2s+1)\theta),
                                                          \label{PhiAjBsCs}
\ee
and $\sum_{j=0}^{\infty} A_j =: A$, say. Here we set
\be
 \tilde{B}_s := \sum_{j=s}^{\infty} B_{j,j-s}, \ \  s=1,2,\ldots  
      \quad
 \tilde{C}_s := \sum_{j=s}^{\infty} C_{j,j-s}, \ \  s=0,1,2,\ldots,
\ee
and also $\tilde{B}_s \equiv \sum_{h=0}^{\infty} B_{s+h,h}$, and
$\tilde{C}_s \equiv \sum_{h=0}^{\infty} C_{s+h,h}$, having set $j-s = h$.

\vskip0.2cm

\vskip0.3cm
\section{The ``vertical force'' due to the potential}   

  Having computed the gravitational potential, we can evaluate the corresponding
``vertical force'', that is that orthogonal to the plane of the two discs. We
have, in the general case, {\em formally},
$$
   -\frac{\p \Phi}{\p z} =
     \sum_{k=0}^{\infty} {-1/2 \choose k} 
 \left[\frac{\p a_1}{\p z} (-1)^k \alpha_1^k + a_1 (-1)^k k \, \alpha_1^{k-1} 
      \frac{\p \alpha_1}{\p z}      \right.
$$
\be
 \left.    + \frac{\p a_2}{\p z} \alpha_2^k + a_2 \, k \, \alpha_2^{k-1}
    \frac{\p \alpha_2}{\p z} \right] \, \cos^k \theta.   \label{z-derivative}
\ee
The convergence of the various series is discussed in Appendix~A.
$$
 \frac{\p a_{\ell}}{\p z} = -G m_{\ell} \, \frac{z}{(R^2 + d_{\ell}^2 + z^2)^{3/2}}
                      = -G z m_{\ell} D_{\ell}^{-3/2},
$$
$$
 \frac{\p \alpha_{\ell}}{\p z} = - \frac{4 R d_{\ell} z}{(R^2 + d_{\ell}^2 + z^2)^2}
                         = - 4 R d_{\ell} z D_{\ell}^{-2},     \ \  \ell =1,2, 
$$
see (\ref{Di}), hence, we obtain, after some algebra,
$$
  \frac{\p}{\p z} \left(a_2 \alpha_2^{2j} + a_1 \alpha_1^{2j} \right)
      = - G z 2^{2j} (4j+1) R^{2j} \left[m_2 d_2^{2j} D_2^{-2j-3/2}
              + m_1 d_1^{2j} D_1^{-2j-3/2}\right],   
$$
$$
  \frac{\p}{\p z} \left(a_2 \alpha_2^{2j+1} - a_1 \alpha_1^{2j+1} \right)
$$
$$
   = - G z 2^{2j+1} (4j+3) R^{2j+1} \left[m_2 d_2^{2j+1} D_2^{-2j-5/2}
              - m_1 d_1^{2j+1} D_1^{-2j-5/2}\right].   
$$
We can then evaluate
\be
  \frac{\p A_j}{\p z} = - \frac{(4j-1)!!}{2^{2j} (j!)^2} (4j+1) G z R^{2j} 
     \left[m_2 d_2^{2j} D_2^{-2j-3/2} + m_1 d_1^{2j} D_1^{-2j-3/2}\right],
                                                                \label{dAj}
\ee
for $j=0,1,2,\ldots$,%
\be
   \frac{\p \tilde{B}_s}{\p z} =
     - 2 G z \sum_{j=s}^{\infty} \frac{(4j+1)!!}{2^{2j} (j-s)! (j+s)!} R^{2j} 
         \left[m_2 d_2^{2j} D_2^{-2j-3/2} + m_1 d_1^{2j} D_1^{-2j-3/2}\right],
                                                                \label{dBs}
\ee
for $s=1,2,3,\ldots$,
$$
   \frac{\p \tilde{C}_s}{\p z} =
      2 G z \sum_{j=s}^{\infty} \frac{(4j+3)!!}{(j-s)! (j+s+1)!} R^{2j+1} 
           \left[m_2 d_2^{2j+1} D_2^{-2j-5/2}    \right.
$$
\be
  \left.   - m_1 d_1^{2j+1} D_1^{-2j-5/2}\right],
                                                               \label{dCs}
\ee
for $s=1,2,3,\ldots$. We conclude with
$$ 
  -\frac{\p \Phi}{\p z} =
      -G z \sum_{j=0}^{\infty} \frac{(4j+1)!!}{2^{2j} (j!)^2} R^{2j} 
     \left[m_2 d_2^{2j} D_2^{-2j-3/2} + m_1 d_1^{2j} D_1^{-2j-3/2}\right]
$$
$$
 - 2 G z \sum_{s=1}^{\infty} \left\{\sum_{j=s}^{\infty} \frac{(4j+1)!!}{2^{2j} (j-s)!
          (j+s)!} R^{2j} \left[m_2 d_2^{2j} D_2^{-2j-3/2}      \right. \right.
$$
$$
 \left. \left.   + m_1 d_1^{2j} D_1^{-2j-3/2} \right] \right\} \cos(2 s \theta)
$$
$$
  + 2 G z \sum_{s=0}^{\infty} \left\{\sum_{j=s}^{\infty} \frac{(4j+3)!!}{2^{2j} (j-s)!
      (j+s+1)!} R^{2j+1} \left[m_2 d_2^{2j+1} D_2^{-2j-5/2}    \right. \right.
$$
\be
   \left. \left.
       - m_1 d_1^{2j+1} D_1^{-2j-5/2} \right] \right\} \cos((2s+1) \theta).
                                                        \label{Phi-AjBsCs}
\ee
The {\em uniform} convergence (with respect to $z$) of all series involved here,
legitimates differentiating with respect to $z$ under the sign of series. This
issue is discussed in Appendix.

\vskip0.1cm

 In view of the smallness of $d_1$, $d_2$, and $z$ with respect to $r$, we have,
approximately, in the general asymmetric case ($m_1 \neq m_2$), observing that
$m_1 d_1 = m_2 d_2$ and recalling that $d_1 + d_2 = d$,
$$
  a_2 \alpha_2 - a_1 \alpha_1 \approx \frac{3 m_1 d_1}{R^2} \, G \,
     \frac{d_1^2 - d_2^2}{R^2},                     
    = \frac{3 m_1 d_1 d G}{R^4} \, (d_1 - d_2),     
$$
$$
  a_2 \alpha_2^2 + a_1 \alpha_1^2 \approx \frac{4 m_1 d_1}{R^3} \, G d \,
   \left[1 - \frac{5}{2 R^2} \left(\frac{d_1^3 + d_2^3}{d} + z^2\right) \right],
$$
hence
$$
    -\Phi = (a_1 + a_2) -\frac{1}{2} [a_2 \alpha_2 - a_1 \alpha_1] \, 
  \cos \theta + \frac{3}{8} [a_2 \alpha_2^2 + a_1 \alpha_1^2] \, \cos^2 \theta
           + \dots
$$
$$
   \approx \frac{m_1 + m_2}{R} \, G
 - \frac{3}{2} \frac{m_1 d_1}{R^2} \, G \, \frac{d_1^2 - d_2^2}{R^2} \cos \theta
$$
$$
     + \frac{3}{2} \frac{m_1 d_1}{R^3} \, G d \, 
   \left[1 - \frac{5}{2 R^2} \left(\frac{d_1^3 + d_2^3}{d} + z^2 \right) \right]
      \, \frac{1 + \cos(2 \theta)}{2}
$$
$$
  = \left\{\frac{m_1 + m_2}{R} \, G + \frac{3}{4} \frac{m_1 d_1}{R^3} \, G d \, 
   [1 - \frac{5}{2 R^2} \left(\frac{d_1^3 + d_2^3}{d} + z^2 \right) \right\}    
$$
$$
 - \frac{3}{2} \frac{m_1 d_1}{R^2} \, G \, \frac{d_1^2 - d_2^2}{R^2} \cos \theta
            + \frac{3}{4} \frac{m_1 d_1}{R^3} \, G d \, 
   \left[1 - \frac{5}{2 R^2} \left(\frac{d_1^3 + d_2^3}{d} + z^2\right) \right]
      \, \cos(2 \theta),
$$
where $\theta := \varphi - \Omega t$. Therefore, 
$$
  -\Phi \approx \left\{ \frac{m_1 + m_2}{R} \, G
         - \frac{3}{4} \, \frac{m_1 d_1}{R^3} \, G d \right\} 
     - \frac{3}{2} \frac{m_1 d_1}{R^4} \, G d \, (d_1 - d_2) \,
           \cos(\varphi - \Omega t) 
$$
\be
  + \frac{3}{4} \, \frac{m_1 d_1}{R^3} \, G d \, \cos(2(\varphi - \Omega t)). 
                                                     \label{approx-asym-pot}
\ee

  In the special case of equal masses (hence, $d_1 = d_2 = d/2$), we have
$$
  -\Phi = 2 a_1 \left\{1 + {-1/2 \choose 2} \alpha_1^2 \cos^2 \theta
      + {-1/2 \choose 4} \alpha_1^4 \cos^4 \theta + \ldots \right\}
$$
$$
  \approx \frac{2 m_1}{R} \, G \, \left\{1 + \frac{3}{16} \, \frac{d^2}{R^2}
         + \frac{3}{16} \, \frac{d^2}{R^2} \, \cos 2 \theta \right\},
$$
i.e.,
\be
    -\Phi \approx \frac{2 m_1}{R} \, G \, \left\{\left(1 + \frac{3}{16} \,
          \frac{d^2}{R^2} \right)
    + \frac{3}{16} \, \frac{d^2}{R^2} \, \cos(2(\varphi - \Omega t))\right\}
                                                  \label{approx-symm-poten}
\ee

  Note that in (\ref{asymm-Phi}), (\ref{symm-Phi}), there will be, at every
order, a new contribution, though smaller and smaller, due to $k$th power of
$\cos^k \theta$. The full formulae could be worked out recalling that
$\cos^k \theta$ could be expressed through Chebyshev polynomials of the first
kind, in fact
$$
     \cos^k \theta =
  \frac{2}{2^k} \sum_{i=0}^{\frac{k-1}{2}} {k \choose i} \, \cos((k - 2i) \theta),
                                        \quad \mbox{for} \ k \ \mbox{odd},
$$
hence  % with $k = 2j+1$
$$
    \cos^{2j+1} \theta =
  \frac{2}{2^{2j+1}} \sum_{i=0}^j {2j+1 \choose i} \, \cos((2j + 1 - 2i) \theta)
$$
\be
    = \frac{1}{2^{2j}} \sum_{i=0}^j {2j+1 \choose i} \, T_{2j+1-2i}(\cos \theta)
                                                      \quad j = 0,1,2, \ldots
\ee
and
$$
      \cos^k \theta = \frac{1}{2^k} {k \choose \frac{k}{2}}
 + \frac{2}{2^k} \sum_{i=0}^{\frac{k}{2}-1} {k \choose i} \, \cos((k - 2i) \theta),
                                       \quad \mbox{for} \ k \ \mbox{even},
$$
hence   
$$
     \cos^{2j} \theta = \frac{1}{2^{2j}} {2j \choose j}
  + \frac{2}{2^{2j}} \sum_{i=0}^{j-1} {2j \choose i} \, \cos((2j - 2i) \theta)
$$
\be
     = \frac{1}{2^{2j}} {2j \choose j}
  + \frac{2}{2^{2j}} \sum_{i=0}^{j-1} {2j \choose i} \, T_{2j-2i}(\cos \theta),
                             \quad j = 1,2,3, \ldots,   %% \sum = 0 if j=0
\ee
while $\cos^0 \theta = T_0(\cos \theta) = 1$, being $\cos n \theta =:
T_n(\cos \theta)$ the $n$th Chebyshev polynomial of the first kind.

\vskip0.3cm
\section{Solution with pseudo-Newtonian Paczy\'{n}ski-Wiita potential}

  It is remarkable that in a number of cases, the pseudo-Newtonian potential
called Paczy\'{n}ski-Wiita (PW, for short) potential
\cite{Paczynski,Abramowicz}, provides very accurate approximations to be used
in place of relativistic quantities. The PW potential corresponding to the mass
$m$ is defined as
$$
          \Phi_{PW} := - \frac{G m}{r - r_G},
$$
where
\be
      r_G = \frac{2 G m}{c^2}                             \label{S-radius}
\ee
is the gravitational or Schwarzchild radius of the body of mass $m$ (defining
the event horizon). In case of two masses, rotating around each other, indeed,
around their common barycenter, the PW potential reads
$$
    \Phi_{PW} = \Phi_{PW}^{(1)} + \Phi_{PW}^{(2)}
             = -\frac{G m_1}{r_1 - r_{G_1}} -\frac{G m_2}{r_2 - r_{G_2}}
$$
We also recall that at distance $r$ from the given mass $m$ equal to $2 r_G$,
the orbits start being unbounded, while at the distance $3 r_G$ there is the
last stable orbit. If we would confining ourselves to distances $r \geq 3 r_G$,
being
$$
  \Phi_{PW} = \frac{G m}{r - r_G} = \frac{G m}{r} \, \frac{1}{1 - \frac{r_G}{r}},
              \equiv \Phi_N \, \frac{1}{1 - \frac{r_G}{r}},
$$
the Newtonian potential $\Phi_N$ will be amplified by a factor $1/(1 - r_G/r)
\leq 3/2$, and at larger distances, this will be less than 3/2.

  Expanding in any case the geometric series factor (convergent for every value
of $r > r_{G_1}, r_{G_2}$, we have
$$
     \Phi_{PW} = \Phi_N^{(1)} \frac{1}{1 - \frac{r_{G_1}}{r_1}}
                + \Phi_N^{(2)} \frac{1}{1 - \frac{r_{G_2}}{r_2}}
$$
$$
 = \Phi_N^{(1)} \left[1 + \sum_{k=1}^{\infty} \left(\frac{r_{G_1}}{r_1}
                  \right)^k\right]
    + \Phi_N^{(2)} \left[1 + \sum_{k=1}^{\infty} \left(\frac{r_{G_2}}{r_2}
                  \right)^k\right]
$$
\be
   = \Phi_N^{(1)} \left[1 + \sum_{k=1}^{\infty} \left(-\frac{2}{c^2}\right)
         \left(\Phi_N^{(1)}\right)^k\right]
   + \Phi_N^{(2)} \left[1 + \sum_{k=1}^{\infty} \left(-\frac{2}{c^2} \right)
           \left(\Phi_N^{(2)}\right)^k\right],              \label{Phi-PW}
\ee
since, by (\ref{S-radius}), 
$$
 \frac{r_{G_{\ell}}}{r_{\ell}} = -\frac{2}{c^2} \, \Phi_N^{(\ell)},   \ \   \ell=1,2.
$$
\vskip0.3cm
\section{The ``vertical force'' due to the potential}

  Having computed the gravitational potential, we can evaluate the corresponding
``vertical force'', that is that orthogonal to the plane of the two discs. We
have, in the general case,
$$
   -\frac{\p \Phi}{\p z} =
     \sum_{k=0}^{\infty} {-1/2 \choose k} 
 \left[\frac{\p a_1}{\p z} (-1)^k \alpha_1^k + a_1 (-1)^k k \, \alpha_1^{k-1} 
      \frac{\p \alpha_1}{\p z}      \right.
$$
\be
 \left.    + \frac{\p a_2}{\p z} \alpha_2^k + a_2 \, k \, \alpha_2^{k-1}
    \frac{\p \alpha_2}{\p z} \right] \, \cos^k \theta.   \label{z-derivative}
\ee
Evaluating
$$
 \frac{\p a_{\ell}}{\p z} = -m_{\ell} G \, \frac{z}{(R^2 + d_{\ell}^2 + z^2)^{3/2}},
               \quad
 \frac{\p \alpha_{\ell}}{\p z} = -\frac{4 R d_{\ell} z}{(R^2 + d_{\ell}^2 + z^2)^2},
                                                           \ \   \ell =1,2, 
$$
we have, confining to the first few terms, after a rather lengthy but
elementary algebra, and recalling that $m_1 d_1 = m_2 d_2$, and that we are
primarily interested in the case $d_i \ll R$, $z \ll R$,
$$
   -\frac{\p \Phi}{\p z} \approx - \frac{G z}{R^3} \left\{\left(m_1 + m_2
  + \frac{15}{4} \, \frac{m_1 d_1 d}{R^2} \right) - \frac{15}{2} \, m_1 d_1
    d \, \frac{d_1 - d_2}{R^3} \, \cos(\varphi - \Omega t)) 
                                              \right.
$$
\be
    \left.   + \frac{15}{4} \, \frac{m_1 d_1 d}{R^2} \,
            \cos(2(\varphi - \Omega t)) \right\},
                                              \label{z-deriv-simple-better}
\ee
being $d = d_1 + d_2$.

\vskip0.1cm

  If, in addition to the previous assumptions, we consider the case of two
equal masses, $m_2 = m_1$ (which implies that $d_1 = d_2 = d/2$), we obtain
\be
   -\frac{\p \Phi}{\p z} \approx - \frac{2 G m_1 z}{R^3} \left\{\left(1
        + \frac{15}{16} \, \frac{d^2}{R^2} \right) 
   + \frac{15}{16} \, \frac{d^2}{R^2} \, \cos(2 (\varphi - \Omega t)) \right\}.
                                                    \label{z-deriv-equal}
\ee
Note that $m_1 d_1^2$ is the {\em moment of inertia} of the mass $m_1$ around
the $z$-axis.

\vskip0.1cm

  In the opposite case of very much different masses, say $m_1 \gg m_2$ (which
entails that $d_1 = m_2 d_2/m_1 \ll d_2$, but still with $d_2 \ll R$, $z \ll
r$), we have instead
$$
   -\frac{\p \Phi}{\p z} \approx - \frac{G m_1 z}{R^3} \left\{\left(1
       + \frac{15}{4} \, \frac{d_1 d_2}{R^2}\right) 
   + \frac{15}{4} \, \frac{d_1 d_2^2}{R^3} \, \cos(\varphi - \Omega t) 
                                                                 \right.
$$
\be
   \left.
  + \frac{15}{8} \frac{d_1 d_2}{R^4} \, \cos(2(\varphi - \Omega t)) \right\}.
                                                 \label{z-deriv-different}
\ee

\vskip0.3cm
\section{The radial component of the gravitational gradient}

  From (\ref{gravit-pot}) we have
$$
  \frac{\p \Phi}{\p R} = \frac{G m_1 \, [R - d_1 \, \cos(\varphi - \Omega t)]}
       {[R^2 + d_1^2 - 2 R d_1 \, \cos(\varphi - \Omega t) + z^2]^{3/2}}
$$
$$
  + \frac{G m_2 \, [R - d_2 \, \cos(\varphi - \Omega t)]}
       {[R^2 + d_2^2 - 2 R d_2 \, \cos(\varphi - \Omega t) + z^2]^{3/2}}.
$$

  From relations (\ref{potential}), (\ref{a-al-ell}) we have
$$
    -\frac{\p \Phi}{\p R}
  = \frac{\p a_1}{\p R} \, \frac{1}{(1- \al_1 \cos \theta)^{1/2}}
     + \frac{a_1}{2} \, \frac{1}{(1- \al_1 \cos \theta)^{3/2}}
     \, \frac{\p \al_1}{\p R}
$$
\be
     + \frac{\p a_2}{\p R} \, \frac{1}{(1- \al_2 \cos \theta)^{1/2}}
     + \frac{a_2}{2} \, \frac{1}{(1- \al_2 \cos \theta)^{3/2}} 
            \, \frac{\p \al_2}{\p R}.
\ee
\vskip0.3cm
\section{The azimuthal component of the gravitational gradient}

  Finally, from (\ref{gravit-pot}) we have
$$
 \frac{\p \Phi}{\p \varphi} = \frac{G m_1 \, R d_1 \, \sin(\varphi - \Omega t)}
       {[R^2 + d_1^2 - 2 R d_1 \, \cos(\varphi - \Omega t) + z^2]^{3/2}}
$$
$$
   + \frac{G m_2 \, R d_2 \, \sin(\varphi - \Omega t)}
       {[R^2 + d_2^2 - 2 R d_2 \, \cos(\varphi - \Omega t) + z^2]^{3/2}}.
$$

  From relations (\ref{potential}), (\ref{a-al-ell}) we have
\be
   -\frac{\p \Phi}{\p \varphi} = \frac{1}{2} \, \sin \theta
       \left[\frac{a_1 \al_1}{(1 - \al_1 \cos \theta)^{3/2}}
     + \frac{a_2 \al_2}{(1 - \al_2 \cos \theta)^{3/2}} \right].
\ee
\section*{{\footnotesize Appendix~A}}
\footnotesize{

  In this Appendix, we discuss the convergence of the various series involved in
the representation of the gravitational potential and its gradient.

\vskip0.1cm
\noindent 1. We first consider the series involving $A_j$. The series
$$
   \sum_{j=0}^{\infty} \frac{\p A_j}{\p z} = \sum_{j=0}^{\infty} {-1/2 \choose 2j}
              {2j \choose j} \frac{1}{2^{2j}}
      \frac{\p }{\p z} \left(a_2 \alpha_2^{2j} + a_1 \alpha_1^{2j}\right)      
$$
\be
   = -G z \sum_{j=0}^{\infty} \frac{(4j+1)!!}{(j!)^2}
      \left[m_2 \left(\frac{R d_2}{2 D_2}\right)^{2j} D_2^{-3/2}
    + m_1 \left(\frac{R d_1}{2 D_1}\right)^{2j} D_1^{-3/2}\right],
                                                           \label{App-SAj}
\ee
converges {\em uniformly} with respect to $z$, which also enters $D_{\ell} :=
R^2 + d_{\ell}^2 + z^2$, $\ell=1,2$. In fact, we have, for either $\ell=1$ or
$\ell=2$ fixed, using Stirling approximation as $j \to \infty$,
$$
  \frac{(4j+1)!!}{(j!)^2} = \frac{(4j+2)!}{2^{2j+1} (2j+1)! (j!)^2}
     \sim \frac{4 \sqrt{2}}{\pi} \, 8^{2j}, 
$$
and hence
$$
   \frac{(4j+1)!!}{(j!)^2} \frac{1}{2^{2j+1}}
      \left(\frac{R d_{\ell}}{2 D_{\ell}}\right)^{2j}
    \sim \frac{2 \sqrt{2}}{\pi} \left(\frac{R d_{\ell}}{2 D_{\ell}}\right)^{2j},
$$
and finally convergence is ensured for $2 R d_{\ell}/D_{\ell} \leq 1$. This is
true for all $z \geq 0$, being
$$
   \frac{2 R d_{\ell}}{D_1} = \frac{2 R d_{\ell}}{R^2 + d_{\ell}^2 + z^ 2}
        \leq \frac{2 R d_{\ell}}{R^2 + d_{\ell}^2} < 1,
$$
for all values of $z \geq 0$, the last strict inequality holds since we can
assume that $d_{\ell} < R$. Therefore, the series in (\ref{App-SAj}) can be
majorized by a convergent numerical (i.e., independent of $z$) series, and
hence the series $\sum_{j=0}^{\infty} A_j$ also converges uniformly with respect
to $z$, and it can thus be differentiated termwise.

\vskip0.2cm
\noindent 2. As for the series containing the even harmonics, we first want to
legitimate doing
$$
       \frac{\p \tilde{B}_s}{\p z} \equiv \frac{\p}{\p z}
          \left(\sum_{j=s}^{\infty} B_{j,j-s}\right)
      = \sum_{j=s}^{\infty} \left(\frac{\p }{\p z} B_{j,j-s}\right).
$$
  In fact, both series
\be
  \sum_{j=0}^{\infty} \frac{(4j+1)!!}{j!}^2 \, \rho^{2j},  \quad
      \sum_{j=s}^{\infty} \frac{(4j+3)!!}{(j-s)!(j+s)!} \, \rho^{2j},   \label{A-s}
\ee
converge uniformly with respect to $\rho := R d_{\ell}/2 D_{\ell}$, since, by the
Stirling approximation formula we have for $j \to \infty$
$$
   \frac{(4j+1)!!}{j!}^2 = \frac{4j+2)!}{2^{2j+1} (2j+1)!} \, \frac{1}{(j!)^2} 
     \sim \frac{(4j+2)(4j+1)(4j)!}{(2j+1)(2j)! 2^{2j+1} (j!)^2}
$$
$$
     \sim 8j \, \frac{(4j)!}{(2j)!} \, \frac{1}{2^{2j+1} (j!)^2}
    \sim \frac{4 \sqrt{2}}{\pi} \, \frac{8^{2j}}{2^{2j+1} (j!)^2)}
           \sim \frac{2 \sqrt{2}}{\pi} \, 4^{2j},
$$
and thus
$$
 \frac{(4j+1)!!}{j!}^2 \, \rho^{2j} \sim \frac{2 \sqrt{2}}{\pi} \, (4 \rho)^{2j}.
$$
This show that the first series in (\ref{A-s}) converges uniformly with respect
to $\sigma := 4 \rho = \frac{2 R d_{\ell}}{D_\ell}$ for $\sigma < 1$. But this is
true since we have
$$
   \frac{2 R d_{\ell}}{D_\ell} = \frac{2 R d_{\ell}}{R^2 + d_\ell^2 + z^2} < 1
$$
for every $ z > 0$ (and $\sigma \leq 1$ for $z \geq 0$). Similarly, 
$$
   \frac{(4j+3)!!}{(j-s)!(j+s=1)!} = \frac{(4j+3)(4j+1)}{(j-s)!(j+s+1)(j+s)!}
         \, \frac{(4j)!}{(2j)!}\, \frac{1}{2^{2j}}
           \sim \frac{8 \sqrt{2}}{\pi} \, 4^{2j},
$$
and thus
$$
     \frac{(4j+3)!!}{(j-s)!(j+s+1)!} \, \rho^{2j}
              \sim \frac{8 \sqrt{2}}{\pi} \, \sigma^{2j},
$$
where again $\sigma := 4 \rho = \frac{2 R d_{\ell}}{D_\ell}$ ($< 1$)as $j \to
\infty$. But more, note that
$$
      \sigma \leq \frac{2 R d_{\ell}}{R^2 + d_{\ell}^2} < 1,
$$
for all $z \geq 0$. Then,
$$
  \sum_{s=0}^{\infty} \sum_{j=s}^{\infty} \frac{(4j+3)!!}{(j-s)!(j+s+1)!}
     \left(\frac{2 R d_{\ell}}{D_\ell}\right)^{2j}
        \leq C \sum_{s=0}^{\infty} \sum_{j=s}^{\infty} \sigma^{2j}
    \leq C \sum_{s=0}^{\infty} \sigma^s \sum_{j=s}^{\infty} \sigma^j
$$
$$
         = \frac{C}{(1-\sigma^2)(1-\sigma)} < \infty,
$$
for some constant, $C$.

  This ultimately shows that differentiating under the sign of series with
respect to $z$ (which enters only in $D_{\ell}$) is permissible, for every $z >
0$.

\vskip0.2cm
\noindent 3. Then, we want to legitimate doing
$$
   \frac{\p }{\p z}\left(\sum_{s=1}^{\infty} \tilde{B}_s
         \, \cos(2 s \, \theta)\right)
    = \sum_{s=1}^{\infty} \left(\frac{\p }{\p z} \tilde{B}_s \right)
            \cos(2 s \, \theta).
$$

  The series    
\be
      \sum_{j=s}^{\infty} \frac{\p B_{j,j-s}}{\p z} = 
  \sum_{j=s}^{\infty} {-1/2 \choose 2j} {2j \choose j-s} \frac{2}{2^{2j}}
    \frac{\p}{\p z}\left((a_2 \alpha_2^{2j} + a_1 \alpha_1^{2j}\right)
                                                      \label{App-Bjjs}
\ee
converges for every fixed $s$, $s \geq 1$ since $\frac{\p \tilde{B}_s}{\p z}$
decreases monotonically to zero as $s \to \infty$. 

  But we can prove more, its uniform convergence with respect to $z$. 
obtain differentiating formally the series in (\ref{App-til-Bs}). This will
ensure that interchanging series and derivative is permissibe as well as the
uniform convergence of the series in (\ref{App-til-Bs}). 

\be
    B_{ji} := {-1/2 \choose 2j} \frac{2}{2^{2j}} {2j \choose i}
      \left[a_2 \alpha_2^{2j} + a_1 \alpha_1^{2j} \right],  \ \  j=1,2,\ldots,
                                            \, i=0,1,\ldots,   \label{Bji}
\ee
\be
    C_{ji} := {-1/2 \choose 2j+1} \frac{1}{2^{2j}}
       {2j+1 \choose i} \left[a_2 \alpha_2^{2j+1} - a_1 \alpha_1^{2j+1} \right],
                      \ \  j,i=0,1,2,\ldots,                    \label{Cji}
\ee
\be
 \tilde{B}_s := \sum_{j=s}^{\infty} B_{j,j-s}, \ \  s=1,2,\ldots  
      \quad
 \tilde{C}_s := \sum_{j=s}^{\infty} C_{j,j-s}, \ \  s=0,1,2,\ldots 
\ee
\vskip0.2cm

\end{document}